# Effective Landau Level Diagram of Bilayer Graphene

Jing Li[1], Yevhen Tupikov[1], Kenji Watanabe[2], Takashi Taniguchi[2], Jun Zhu[1,3*]

[1]Department of Physics, The Pennsylvania State University, University Park, Pennsylvania 16802, USA.

[2]National Institute for Material Science, 1-1 Namiki, Tsukuba 305-0044, Japan.

[3]Center for 2-Dimensional and Layered Materials, The Pennsylvania State University, University Park, Pennsylvania 16802, USA.

*Correspondence to: jzhu@phys.psu.edu (J. Zhu)

**Abstract**

The $E = 0$ octet of bilayer graphene in the filling factor range of $-4 < \nu < 4$ is a fertile playground for many-body phenomena, yet a Landau level diagram is missing due to strong interactions and competing quantum degrees of freedom. We combine measurements and modeling to construct an empirical and quantitative spectrum. The single-particlelike diagram incorporates interaction effects effectively and provides a unified framework to understand the occupation sequence, gap energies and phase transitions observed in the octet. It serves as a new starting point for more sophisticated calculations and experiments.

Bilayer graphene provides a fascinating platform to explore potentially new phenomena in the quantum Hall regime of a two-dimensional electron gas (2DEG). The existence of two spins, two valley indices K and K′, and two isospins corresponding to the $n$ = 0 and 1 orbital wave functions results in an eight-fold degeneracy of the single-particle $E$ = 0 Landau level (LL) in a perpendicular magnetic field $B$ [1,2]. This SU (8) phase space provides ample opportunities for the emergence of broken-symmetry many-body ground states [3-15]. The application of a transverse electric field $E$ drives valley polarization through their respective occupancy of the two constituent layers [1,2]. Coulomb exchange interactions, on the other hand, enhance spin ordering and promote isospin doublets [11,15,16]. As a result of these intricate competitions, the $E$ = 0 octet of bilayer graphene (integer filling factor range -4 < $\nu$ < 4) exhibits a far richer phase diagram than their semiconductor counterparts. Experiments have uncovered 4, 3, 2, 1 coincidence points for filing factors $\nu$ = 0, ±1, ±2 and ±3 respectively, where the crossing of two LLs leads to the closing of the gap and signals the phase transition of the ground state from one order to another [13,15-18]. Their appearance provides key information to the energetics of the LLs and the nature of the ground states involved. Indeed, coincidence studies on semiconducting 2DEGs are used to probe the magnetization of quantum Hall states [19] and measure the many-body enhanced spin susceptibility [20]. In bilayer graphene, the valley and isospin degrees of freedom increase the number of potential many-body coherent ground states. Furthermore, the impact of actively controlling these degrees of freedom became evident in the recent observations of fractional and even-denominator fractional quantum Hall effects [17,21-25].

A good starting point of exploring this rich landscape would be a single-particle, or single-particlelike LL diagram, upon which interaction effects can be elucidated perturbatively. Indeed, even in the inherently strongly interacting fractional quantum Hall effect, effective single-particle models, e.g. the composite fermion model [26], can capture the bulk of the interaction effects and provide conceptually simple and elegant ways to understand complex many-body phenomena. In bilayer graphene, a LL diagram that provides a basis to interpret and reconcile the large amount of experimental findings to date has yet to emerge. Predictions of tight-binding models with Hartree-Fock approximations [2,14,27-30] are not able to fully account for experimental observations [16].

Here we have taken an empirical approach to construct an effective single-particlelike LL diagram of bilayer graphene subject to perpendicular magnetic and electric fields. This effective LL diagram provides a unified framework to interpret existing experiments. It can *quantitatively* reproduce the observed coincidence conditions of $\nu$ = 0 and ±1 and account for the widely varying literature reports on the gap energies at $\nu$ = ±1, ±2 and ±3. The diagram produces five filling sequences of the LLs from $\nu$ = -4 to +4, in excellent agreement with experiment [16]. An expression for the energy splitting between the $n$ = 0 and 1 orbitals $E_{10}$ is obtained.

All seven devices reported in this letter are dual-gated, with the bilayer sheet sandwiched between two h-BN dielectric layers. The fabrication procedures and characteristics of the devices can be found in Section 1 of Ref. [31]. The first important energy scale of our diagram is the perpendicular displacement field $D$ induced interlayer potential difference $\Delta(D)$ at $B$ = 0. We determine $\Delta(D)$ using thermally activated transcript measurements. For $D < 800$ mV/nm, $\Delta(D)$ is well approximated by $\Delta$ (meV) = $0.13 D$ (mV/nm) (See Section 2 of the Ref. [31]).

When a perpendicular magnetic field $B$ is applied, the gapped bands of bilayer graphene evolve into discrete LLs. In a two-band tight-binding model when $D$ is large, the $\nu = 0$ gap is approximately give by $\Delta(D)$, which represents the energy splitting between the $n = 0$ orbitals in K and K' valleys, i.e. between $|+,0\rangle$ and $|-,0\rangle$. The splitting between the $n = 1$ orbitals, i.e. $|+,1\rangle$ and $|-,1\rangle$, is slightly smaller due to wave function distributions [2]. This model predicts a LL sequence of $|+,0\rangle$, $|+,1\rangle$, $|-,1\rangle$, $|-,0\rangle$. The effect of the electron-hole asymmetry, however, produces a large positive correction to the energies of the $|\pm,1\rangle$ states and modifies the sequence to $|+,1\rangle$, $|+,0\rangle$, $|-,1\rangle$, $|-,0\rangle$ [14]. The correction $E_{10}$(meV) $= E_1 - E_0 = (\gamma_1 \gamma_4 / \gamma_0) B = 0.48 B$(T), where $\gamma_0$, $\gamma_1$, $\gamma_4$ are the Slonczewski-Weiss-McClure hopping parameters [32], is much larger than the bare Zeeman energy $\Delta_z$(meV) = 0.11 $B$ (T). Thus spin doublets, e.g. $|0,\uparrow\rangle$ followed by $|0,\downarrow\rangle$, are favored in this model [14]. The recent experimental observations of Hunt et al, however, point to the formation of closely spaced orbital doublets at large $D$ [16]. Large exchange corrections presumably play an important role[16], although the effect of trigonal warping has yet to be examined carefully [33]. Beyond tight-binding models, many-body effects are expected to modify the LL gap energies with terms linear in $B$ [10,11].

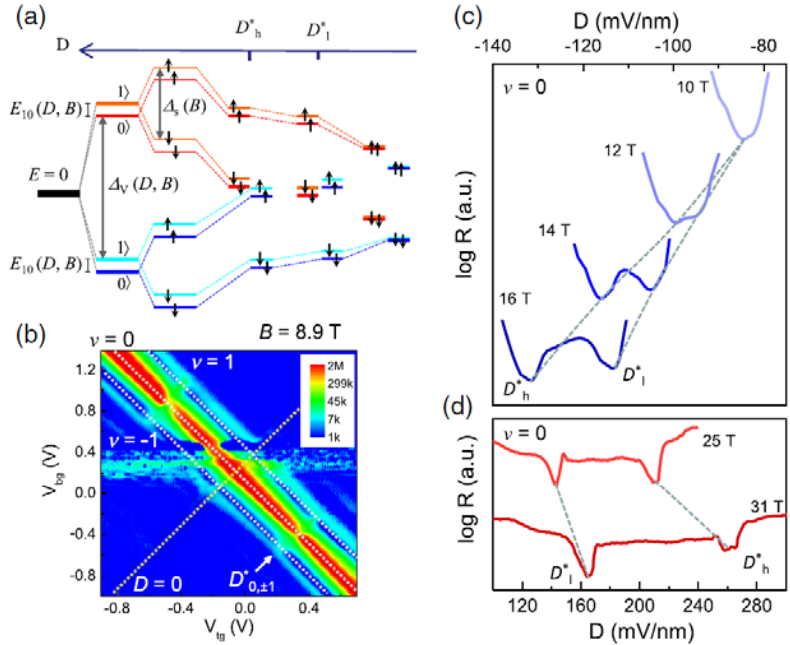

FIG.1. (a) An effective LL diagram for the $E = 0$ octet of bilayer graphene at a fixed magnetic field. Red, orange, blue, cyan colors denote $|+, 0\rangle$, $|+, 1\rangle$, $|-, 0\rangle$ and $|-, 1\rangle$ states respectively, following the color scheme of Ref. [16]. Illustrated are four scenarios corresponding to large $D$, the coincidence fields $D^*_h$ and $D^*_l$ of $\nu = 0$ and small $D$. (b) A color map of $R_{xx}$ ($V_{tg}$, $V_{bg}$) in device 6 at $B = 8.9$ T. Dashed lines mark the constant filling factors $\nu = 0, \pm 1$ and $D = 0$. The arrow points to the positive coincidence fields of $\nu = 0, \pm 1$. They are not distinguishable at this field. Disturbance observed in the range of $0 < V_{bg} < 0.4$ V is due to contact problems. (c) and (d), $R_{xx}$ ($D$) obtained at $\nu = 0$ at selected $B$-fields from 10 - 16 T in device 24 (c), and at $B = 25$ T and 31 T in device 6 (d). The dashed lines are guides to the eye for $D^*_h$ and $D^*_l$. Both are symmetric about 0. Only one direction is shown for each device.

Starting from the prior knowledge, we have constructed an effective single-particlelike LL diagram of bilayer graphene, which is shown in Fig. 1(a). Three energy scales are introduced. We postulate that the valley gap $\Delta_v$ $(D, B)$ between the $|+,0\rangle$ and $|-,1\rangle$ states takes on the form of $\Delta_v$ $(D, B) = \Delta(D) + \alpha B$ and the exchange-enhanced spin gap $\Delta_s$ $(B)$ between states of opposite spin polarizations takes on the form of $\Delta_s$ $(B) = \beta B$. The magnitude and functional form of $E_{10}$ $(D, B)$, which yields the gaps at $\nu = \pm 1$ and $\pm 3$, is to be determined by experiments. The filling sequence in the large $D$ limit is fixed by experiments [16]. The gap at $\nu = 0$ is given by $\Delta_0 \approx |\Delta_v (D, B) - \Delta_s (B)|$ and transitions from a valley gap at large $D$ to a spin gap at small $D$, which is generally consistent with experimental findings, although the spin polarized ground state of $\nu = 0$ only appears in a large in-plane magnetic field [9,11,13]. From large $D$ to small $D$, the gap of $\nu = 0$ closes twice, at $D^*_h$ and $D^*_l$ respectively. The larger $D^*_h$ corresponds to $E_{+,0,\downarrow} = \frac{1}{2}\Delta_v - \frac{1}{2}\Delta_s + \frac{1}{2}E_{10} = E_{-,1,\uparrow} = -\frac{1}{2}\Delta_v + \frac{1}{2}\Delta_s + \frac{1}{2}E_{10}$ i.e. $\Delta_v$ $(D^*_h) = \Delta_s$ $(B)$ or $\Delta (D^*_h) = (\beta - \alpha)B$ while a smaller $D^*_l$ corresponds to $E_{+,1,\downarrow} = \frac{1}{2}\Delta_v - \frac{1}{2}\Delta_s + \frac{3}{2}E_{10} = E_{-,0,\uparrow} = -\frac{1}{2}\Delta_v + \frac{1}{2}\Delta_s - \frac{1}{2}E_{10}$, i.e. $\Delta (D^*_l) + 2E_{10} (D^*_l, B) = (\beta - \alpha)B$ [34]. They are distinguishable so long as $E_{10}$ can be resolved experimentally.

Figures 1(b) - (d) present our measurements of $D^*_h$ and $D^*_l$ for $\nu = 0$. Figure 1(b) shows a colored map of magnetoresistance $R_{xx}$ as a function of the top and bottom gate voltages $V_{tg}$ and $V_{bg}$ in device 06 at $B = 8.9$ T. Lines corresponding to constant filling factors $\nu = 0, \pm 1$ and $D = 0$ are marked in the plot. We sweep the top and bottom gates in a synchronized fashion to follow a line of constant $\nu$ and measure $R_{xx}$ $(D)$. Similar to previous studies, a dip (peak) in $R_{xx}$ $(D)$ is identified as the coincidence field $D^*_0$ ($D^*_{\pm 1}$) for $\nu = 0$ ($\pm 1$) [13,16-18]. $D^*$ is symmetric about 0 and the positive $D^*_{0, \pm 1}$ is marked in Fig. 1(a). In Fig. (c), we plot a few examples of $R_{xx}$ $(D)$ at fixed $B$-fields from 10 - 16 T in device 24. A double-dip structure starts to appear at $B \sim 12$ T and the difference between $D^*_h$ and $D^*_l$ rapidly increases with $B$. Higher field data up to 31 T obtained on device 06 are shown in Fig. 1(d).

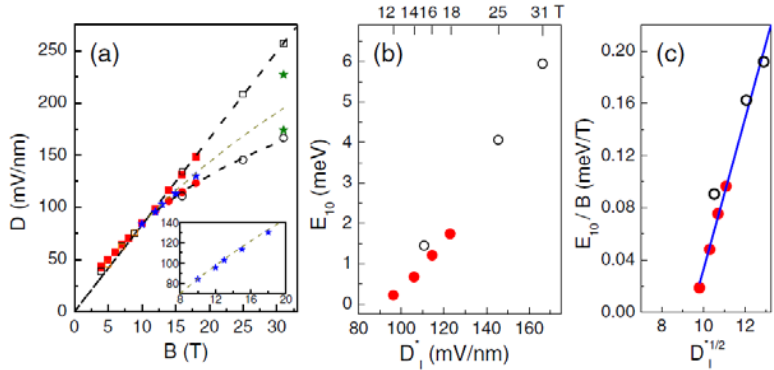

FIG.2. (a) The measured $\nu = 0$ coincidence field $D^*_h$ (squares) and $D^*_l$ (circles) vs. $B$ in devices 24 (red symbols), 06 (black symbols), and 34 (orange symbols). Blue stars plot $D^*_{-1}$ obtained in device 43. Olive stars are data read from Ref. [16] for $D^*_{+1}$ (upper point) and $D^*_{-1}$ (lower point) at $B = 31$ T. Black dashed lines plot $D^*_h = 8.3B$ (the upper branch) and $D^*_l$ (the lower branch) obtained from our diagram. The dark yellow dashed line plots the predicted $D^*_{\pm 1}$. The inset plots the blue stars and the dark yellow dashed line again for clarity. (b) $E_{10}$ vs. $D^*_l$. The top axis marks the corresponding $B$ field. (c) $E_{10}/B$ vs. $\sqrt{D^*_l}$. The blue

line represents a linear fit in the form of $E_{10}/B = 0.058(\sqrt{D_l^*} - 9.43)$. Symbols in (b) and (c) follow the notation of (a).

Figure 2(a) summarizes results of $D^*$, $D^*_h$ and $D^*_l$ obtained from 4 devices. Above $B \sim 7$ T, $D^*_h$ ($D^*$ at low field) exhibits a remarkably linear dependence on $B$, with a slope of 8.3 mV/nm/T. (Considerable deviation of $D^*$ from the line is observed at $B < 5$ T and not discussed in this work.) Both the linear trend and the slope are in good agreement with measurements obtained by other groups on h-BN encapsulated bilayers [13,16-18]. The linear dependence of $D^*_h(B)$, together with $\Delta(D) = 0.13D$, leads to $\beta - \alpha = 1.1$ meV/T.

The appearance of $D^*_l$ at $B > 12$ T enables us to determine the magnitude of $E_{10}$ ($D$, $B$). Figure 2(b) plots $E_{10}$ obtained from devices 06 and 24. $E_{10}$ increases rapidly from 0.2 meV at $D = 96$ mV/nm ($B = 12$ T) to 6.0 meV at $D = 167$ mV/nm ($B = 31$ T). The coincidence studies alone are not sufficient to determine the role of the electric and magnetic fields independently in $E_{10}$. Since the $\nu = 1$ gap is given by $E_{10}$ and has been shown to be approximately linear in $B$ in the literature [22,35-37], we further assume that $E_{10}(D, B)/B$ is a pure function of $D$ and its functional form can then be obtained from data, as shown in Fig. 2(c). $E_{10}/B$ is a strongly non-linear function of $D$ and only rises sharply after a large threshold of $D$-field is reached. We obtain $E_{10}/B = 0.058(\sqrt{D(\text{mV/nm})} - 9.43)$ in the regime of $D \geq 96$ mV/nm, with the choice of the $\sqrt{D}$ form motivated by the linear fit obtained.

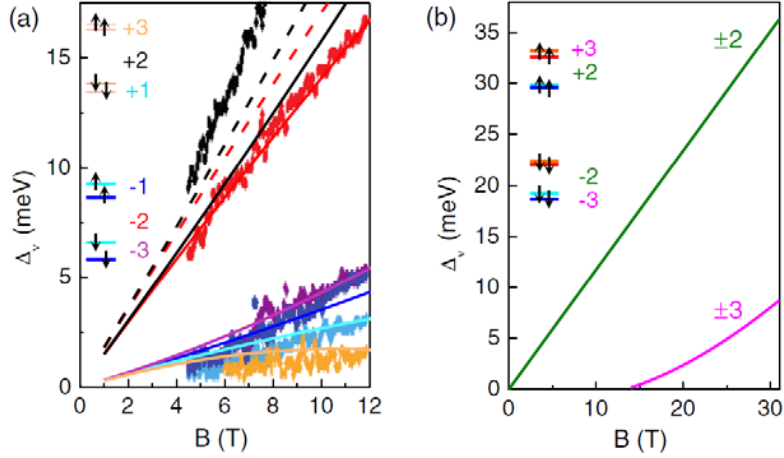

FIG.3. (a) Measured (Ref.[22]) and calculated (smooth curves) gap energies $\Delta_\nu$ at $\nu = \pm 1, \pm 2,$ and $\pm 3$. The measured curves and the filling factor labels in the inset are color coordinated. All calculations use $D_0 = 220$ mV/nm with the field line pointing downwards. The solid red and black curves correspond to $\beta = 1.7$ meV/T. The dashed red and black curves correspond to $\beta = 2.1$ meV/T. The measured $\Delta_{-2}$ has a slope of 1.4 meV/T. (b) $\Delta_{\pm 2}$ and $\Delta_{\pm 3}$ calculated with $D_0 = 0$ and $\beta = 1.7$, $\alpha = 0.6$ meV/T. $\Delta_{\pm 1}$ is below the limit of our calculation. $\Delta_{\pm 2}$ has a slope of 1.2 meV/T. The insets of (a) and (b) illustrate the LL sequence corresponding to each scenario respectively.

The knowledge of $E_{10}(D, B)$ sheds considerable light on the widely varying reports of the LL gap energies in the literature [15,18,22,35-37]. In Fig. 3(a), we compare the systematic measurements of Kou et al. [22] with calculations produced by our effective model. In the singly-gated sample used in Ref. [22], the $D$-field grows with carrier density $n$, which translates into a filling factor-dependent quantity $D(\nu) = 2.2\nu B$ (mV/nm). In addition, the sample may have unintentional chemical doping, the compensation of which results in a finite $D_0$ at $\nu = 0$. Together, the sample experiences $D = D_0 + 2.2\nu B$.

The large band gap of ~ 25 meV at $B = 0$ observed in Ref. [22] suggests a large $D_0$. As shown in Fig. 3(a), with a single fitting parameter of $D_0 = 220$ mV/nm, our calculated $\Delta_v$'s can capture the size and order of the measurements at $\nu = \pm 1$ and $\pm 3$ very well, attesting to the strength of our model. Moreover, values of $\beta = 1.7$-$2.1$ meV/T put the calculated gaps of $\nu = \pm 2$ in the vicinity of the measured data. Here, the measurements show a larger asymmetry between $\nu = +2$ and $-2$ than our simulations would suggest. One possibility of this discrepancy can be due to the different many-body screening at $\nu = +2$ and $-2$, which requires more sophisticated calculation to capture.

An estimated $\beta = 1.7$ meV/T, together with $\beta - \alpha = 1.1$ meV/T obtained earlier, leads to an $\alpha = 0.6$ meV/T and the quantitative knowledge of all three energy scales $\Delta_v (D, B)$, $\Delta_s (B)$ and $E_{10} (D, B)$ used in our effective LL diagram. We discuss a number of insights obtained by examining the diagram in a wide range of $D$- and $B$-fields. First of all, it is instructive to compare the large-$D$ scenario represented in Fig. 3(a) with that of a small $D$. Figure 3(b) plots the calculated $\Delta_v$'s at $\nu = \pm 2$ and $\pm 3$ with $D_0 = 0$ corresponding to no unintentional doping. $\Delta_{\pm 1}$ is too small to be calculated accurately. The vanishing gap of $\nu = \pm 1$ and $\pm 3$ at small $D$ and the gap-enhancing effect of the $D$-field, corroborates many experiments in the literature [5,18,22,35,36] and is also supported by our data (See Fig. S3 (a) in Ref. [31]).

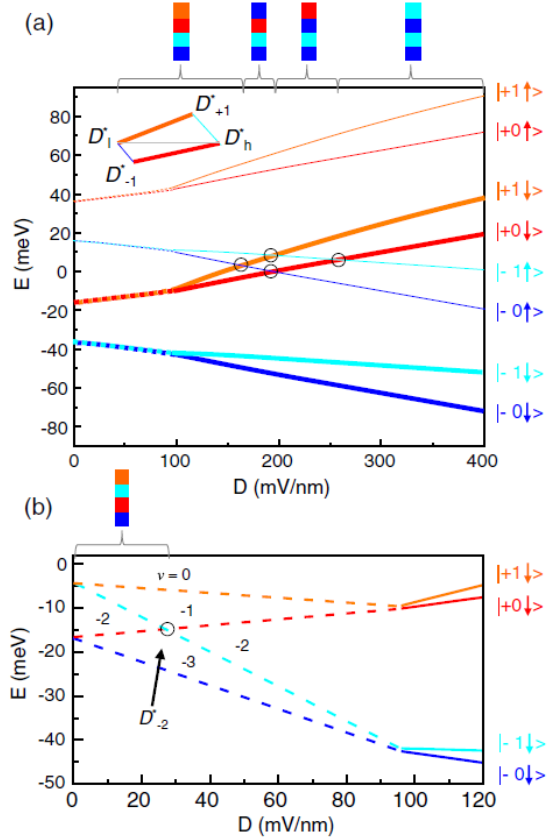

FIG.4. (a) The effective LL diagram for the $E = 0$ octet calculated for a fixed $B = 31$ T and as a function of $D$. The eight states are labeled on the right. Color bars illustrate the filling sequences from $\nu = -4$ to 0. The four coincidence fields for $\nu = 0, \pm 1$ are circled. The inset illustrates the order observed in Ref. [16]. (b) A qualitative sketch of the LL diagram at low $D$-field in the vicinity of $\nu = -2$.

The behavior of $v = \pm 2$ is markedly different. As the insets of Figs. 3(a) and (b) illustrate, the nature of the $v = \pm 2$ gap changes from a spin-splitting origin at large $D$ to a valley-splitting origin at small $D$. The transition occurs near $D^*$ of $v = 0$ (See Fig. 4(a)). The magnitude of the gap, however, has slopes of 1.4 and 1.2 meV/T respectively in the two regimes. This transition is thus difficult to detect based on gap measurements alone. Indeed, measurements of $\Delta_{\pm 2}$ in the literature have all reported slopes of 1-1.4 meV/T [15,22,35,36], in excellent agreement with the predictions of our model. The large gaps at $v = \pm 2$, in both scenarios, result from many-body enhancement and are effectively represented in our single-particlelike diagram.

Figure 4(a) plots a full diagram of the $E = 0$ octet, calculated by fixing $B = 31$ T and varying the $D$-field. Four coincidences are seen. The calculated $D^*_h$ and $D^*_l$ for the $v = 0$ state are plotted as black dashed lines in Fig. 2(a) and match data well. This is expected since we have reverse-engineered our diagram based on these observations. In addition, the diagram predicts the closing of the $v = \pm 1$ gaps at $D^*_{\pm 1}$, the value of which is calculated and plotted in Fig. 2(a) as a dark yellow dashed line. Also plotted there are our measurements for $D^*_{-1}$ (blue stars) obtained in device 43 using maps similar to that shown in Fig. 1(a), and data points obtained by Hunt et al [16] at $B = 31$ T (olive stars). The calculated $D^*_{\pm 1}$ is e-h symmetric and captures the average of the measured $D^*_{+1}$ and $D^*_{-1}$ very well. However both our data and that of Ref. [16] systematically deviate from the calculated $D^*_{\pm 1}$, with $D^*_{-1}$ tending towards $D^*_l$ and $D^*_{+1}$ tending towards $D^*_h$. This intrinsic asymmetry between $v = \pm 1$ is schematically illustrated in the inset of Fig. 4(a). They point to $v$-dependent many-body screening effects missing in model. Similarly, $v$-dependent phase transition lines within each LL [16] is also not captured.

In our model, $E_{10}$ vanishes in the vicinity of $D \sim 100$ mV/nm. A negligible $E_{10}$ down to $D = 0$ would lead to LLs shown in Fig. 4(a), where $v = \pm 2$ remain valley-split in nature. Experimentally, $v = \pm 2$ undergoes another transition at small $D$-field, possibly to an isospin polarized ground state[15], as illustrated in Fig. 4(b). The coincidence field $D^*_{-2}$ occurs at $\sim 27$ mV/nm at 31 T [16] and exhibits a rough slope of $\sim 0.9$ mV/nm/T at lower field [15,17]. The scenario sketched in Fig. 4(b) is consistent with the reported filling sequence below $D^*_{-2}$ [16], and the observations of vanishing $\Delta_{\pm 1}$ and $\Delta_{\pm 3}$ at $D = 0$ [16,17]. A more quantitative understanding of this part of the LL diagram would require careful, direct measurements of $E_{10}$ at low $D$-fields. An accurate knowledge of $E_{10}$ would also aid the understanding and control of even-denominator fractional quantum Hall states in bilayer graphene, which so far only occur in the $n = 1$ orbitals [21,23,24].

The diagrams shown in Figs. 4(a) and (b) together reproduce the five $D$-dependent filling sequences of the $E = 0$ octet, which are illustrated above the graphs [16]. The agreement is quite remarkable and attests to the validity of the effective single-particle-like approach in capturing many features of the complex many-body system.

A qualitative failure of our model occurs at $v = 0$ in low $D$-field, where a spin ferromagnet is predicted while experiments point to a canted antiferromagnet with spin-valley coherence [10,11,13]. This single-particlelike diagram is also likely to fail near crossing points, where quantum Hall ferromagnets coherent in more than one degree of freedom may occur [15,37]. We hope that our model provides a skeleton, upon which more sophisticated theoretical tools and measurements can be built to illuminate the rich quantum Hall physics bilayer graphene has to offer.

In summary, we constructed an empirical LL diagram for the $E = 0$ octet of bilayer graphene in the presence of perpendicular magnetic and electric fields. This diagram offers a unified, intuitive framework to interpret many experimental findings to date, complete with quantitative energy scales. We hope that it serves as a good base to launch future experiments and calculations.

**Acknowledgment**

Work at Penn State is supported by the NSF through NSF-DMR-1506212. Work at NIMS is supported by the Elemental Strategy Initiative conducted by the MEXT, Japan and JSPS KAKENHI Grant Number JP15K21722. Part of this work was performed at the NHMFL, which was supported by the NSF through NSF-DMR-1157490 and the State of Florida. We thank Herbert Fertig for helpful comments and Jan Jaroszynski of the NHMFL for experimental assistance.

# Supplementary material for

## Effective Landau Level Diagram of Bilayer Graphene

Jing Li[1], Yevhen Tupikov[1], Kenji Watanabe[2], Takashi Taniguchi[2], Jun Zhu[1,3*]

[1]Department of Physics, The Pennsylvania State University, University Park, Pennsylvania 16802, USA.

[2]National Institute for Material Science, 1-1 Namiki, Tsukuba 305-0044, Japan.

[3]Center for 2-Dimensional and Layered Materials, The Pennsylvania State University, University Park, Pennsylvania 16802, USA.

*Correspondence to: jzhu@phys.psu.edu (J. Zhu)

**Online Supplementary Information Content**

1. Device fabrication and characteristics
2. Temperature dependence at the charge neutrality point
3. Landau level diagrams

# 1. Device fabrication and characteristics

Data from seven *h*-BN encapsulated, dual-gated devices are used in this work. Devices 01 and 24 are fabricated using a van der Waals dry transfer plus side contacts method introduced by Wang et al [1]. They are two-terminal devices. Devices 06, 23L, 23R, 34 and 43 are fabricated using a layer-by-layer transfer method [2]. They are multi-terminal Hall bar devices. Devices 06 and 34 were used in our previous work [3] with the fabrication procedure and the methods to characterize and use the top and bottom gates given in detail in the supporting information of the paper (Devices 06 and 34 are devices 1 and 2 respectively in Ref. [3]). The mobility of the devices ranges from 20, 000 to 100, 000 cm$^2$V$^{-1}$s$^{-1}$. Figure S1 shows a semi-log plot of $R_{CNP}$ ($D$) for five of our devices. $R_{CNP}$ grows nearly exponentially with $D$ in all our devices, with higher slope corresponding to higher device quality.

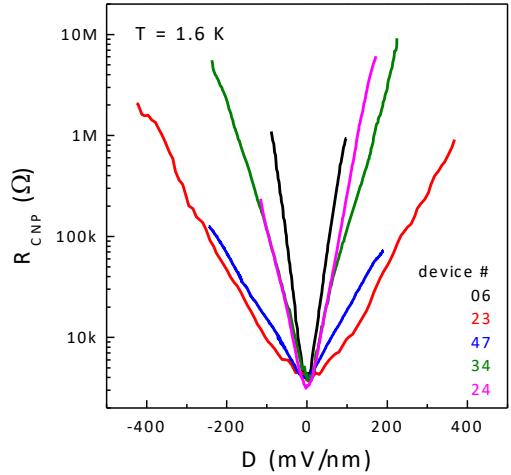

Figure S1. $R_{CNP}$ vs. $D$ in a semi-log plot for devices 06, 23, 47, 34 and 24 as labeled in the plot. Devices 01 is similar to device 24. Device 43 is similar to device 47. A contact resistance of 700 Ω is subtracted from the two-terminal resistance of device 24. The curves are shifted in the vertical direction to facilitate comparison.

# 2. Temperature dependence at the charge neutrality point

We measure *Δ* as a function of the applied displacement electric field *D* using thermally activated transport at the charge neutrality point (CNP). *D* is defined as

$$D = \frac{1}{2}\left(\frac{\varepsilon}{d_b}(V_{bg} - V_{bg0}) - \frac{\varepsilon}{d_t}(V_{tg} - V_{tg0})\right)$$

[4] [5], where $\varepsilon = 3$ is the dielectric constant of h-BN, $d_t$($d_b$) the top(bottom) h-BN thickness and $V_{tg}$ ($V_{bg}$) the applied top(bottom) gate voltage. $V_{tg0}$ and $V_{bg0}$ are offsets given by unintentional chemical doping. Figure S2(a) shows the temperature dependence of the resistance $R_{CNP}$ (*T*) at selected *D* fields in device 23L in an

Arrhenius plot. We fit our data to $R_{CNP} \sim \exp(\Delta/2k_BT)$ and extract $\Delta$ using the linear fits in Fig. S2(a). Such analysis was performed for a range of negative and positive $D$-fields in devices 23L and 23R. The resulting $\Delta(D)$ from different data sets overlap very well, as shown in Fig. S2(b). In Fig. S2(c), we show one set of $\Delta(D)$ (solid squares), together with previous results obtained using optical spectroscopy by Zhang et al [4] (open squares). They agree well with each other. The solid red line in Fig. S2(c) represents a polynomial fit to our data. At $D < 800$ mV/nm, which covers all of the quantum Hall studies in the literature, $\Delta(D)$ is well approximated by $\Delta$ (meV) = 0.13$D$ (mV/nm). This implies an effective dielectric constant of $\varepsilon = 2.6$ for the interlayer screening of bilayer.

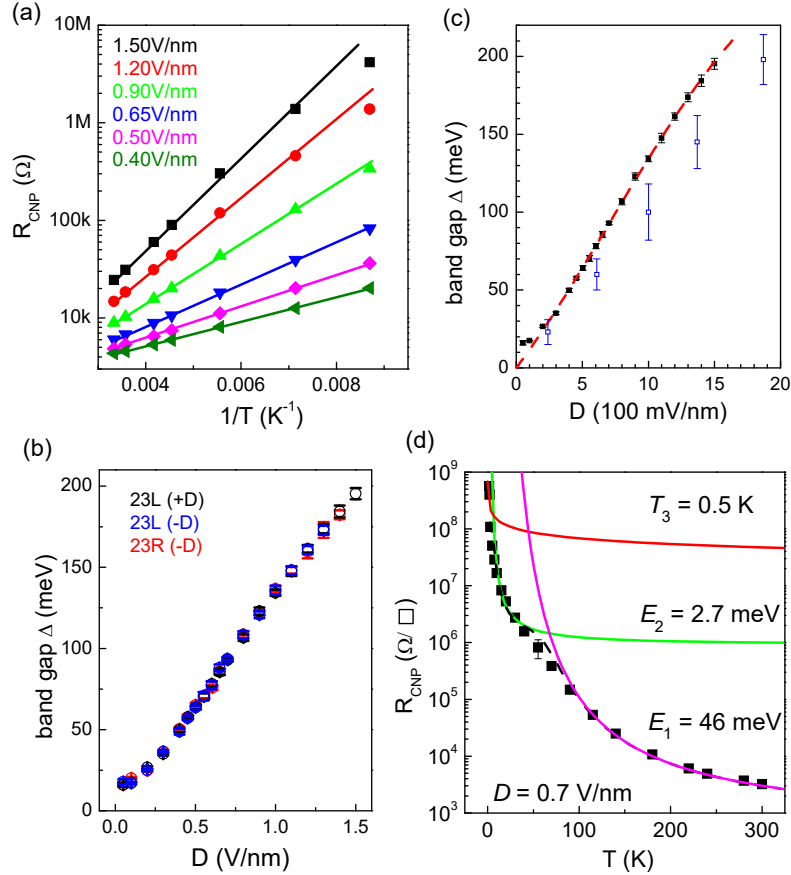

Figure S2. (a) The CNP resistance $R_{CNP}$ vs. $1/T$ in a semi-log plot at selected $D$-fields as labeled in the plot. Solid lines are linear fits to data. From device 23L. (b) Extracted $\Delta(D)$ from three sets of data as labeled in the graph. (c) The extracted band gap $\Delta$ as a function of $D$. From device 23L (solid symbols). The data is well described by $\Delta(D) = 0.113D + 4 \times 10^{-5}D^2 - 1.8 \times 10^{-8}D^3$ in the entire range (red dashed curve). Data below $D = 800$ mV/nm can also be approximated by $\Delta(D) = 0.13D$. Open symbols plot $\Delta(D)$ read from Ref. [4]. (d) Sheet resistance $R_{CNP}(T)$ in a semi-log plot at $D = 0.7$ V/nm. The colored fits represent the three terms in Eq. (S1) with the fitting parameters written next to the curves. The sum of the three is shown as a black dashed line. From device 23L.

As the temperature decreases, $R_{CNP}(T)$ deviates from exp ($\Delta/2k_BT$). We observe behaviors similar to early observations made on oxide-supported devices [5], only with smaller disorder scales due to the clean $h$-BN. Figure S2(d) plots $R_{CNP}(T)$ at $D = 0.7$ V/nm in device 23L, which covers the temperature range of 1.6 - 300 K. The black dashed line is a fit to data using Eq. (S1), which includes three contributions originating from thermally activated transport to the band edge at high temperature (80 K < T < 300 K, magenta curve), nearest neighbor hopping at intermediate temperature (5 K < T < 80 K, green curve) and variable range hopping below 5 K (red curve). A detailed discussion of the three mechanisms can be found in our earlier work [5].

$$\frac{1}{R_{CNP}(T)} = \frac{1}{R_1}\exp(-E_1/k_BT) + \frac{1}{R_2}\exp(-E_2/k_BT) + \frac{1}{R_3}\exp\left[-(T_3/T)^{1/3}\right] \quad .$$

(S1)

The fitting parameters are $E_1 = \Delta/2 = 46$ meV, $E_2 = 2.7$ meV and $T_3 = 0.5$ K.

## 3. Landau level diagrams

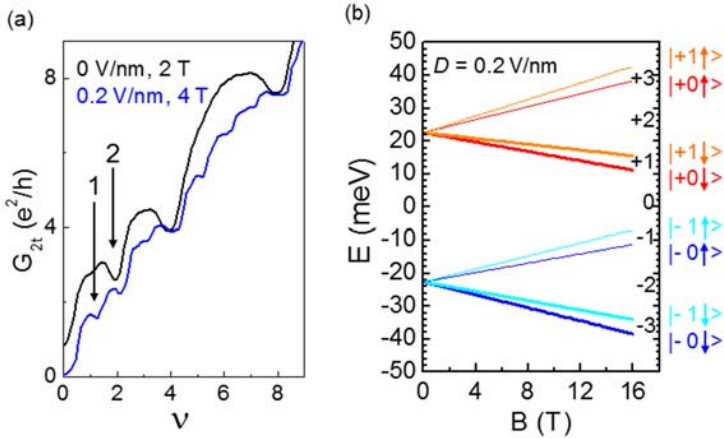

Figure S3. (a) Two-terminal conductance $G_{2t}$ as a function of filling factor $\nu$ at fixed $D$ and $B$ fields as labeled in the plot. Here the magnetic field is fixed and synchronized gate sweeps are used to sweep $\nu$ along constant $D$ lines. From device 01. (b) A calculated Landau level diagram at $D = 0.2$ V/nm showing a measurable gap at $\nu = 1$ at several Teslas.

Figure S3(a) plots the two-terminal conductance $G_{2t}$ of device 01 as a function of the filling factor $\nu$ at $B$- and $D$-fields as labeled in the graph. The $\nu = 2$ state is well developed at $B = 2$ T at $D = 0$ V/nm while the $\nu = 1$ state is only visible at large $D$-field.

These trends supports the Landau level diagram discussed in the text. In Fig. S3(b), we plot the calculated energy diagram of the $E = 0$ octet at $D = 0.2$ V/nm. At this $D$-field, $E_{10}$ is a few meV at several Teslas, which is consistent with the appearance of $v = 1$ in Fig. S3(a).

**Supplementary Reference**